\documentclass[prb,preprint]{revtex4-1}
\usepackage{amsmath}  % needed for \tfrac, \bmatrix, etc.
\usepackage{amsfonts} % needed for bold Greek, Fraktur, and blackboard bold
\usepackage{graphicx} % needed for figures

\begin{document}

\title{Symmetry-Breaking Transition and Spectral Singularity in Coupled $\mathcal{PT}$-Symmetric Quantum Potentials}

\author{Yu Jiang}
\email{jiang@xanum.uam.mx}

\affiliation{Departamento de Fisica, Universidad Autonoma Metropolitana - Iztapalapa, Mexico City, Mexico}

\date{\today}

\begin{abstract}
We study the scattering properties of $N$ identical one-dimensional localized $\mathcal{PT}$-symmetric potentials, connected in series as well as in parallel. We derive a general transfer matrix formalism for parallel coupled quantum scatterers, and apply that theory to demonstrate that the spectral singularities and $\mathcal{PT}$-symmetric transitions of single scattering cells  may be observed in coupled systems, at the same or distinct values of the critical parameters, depending on the connection modes under which the scattering objects are coupled. We analyse the influences of the connection configuration on the related transport properties such as spectral singularities and anisotropic transmission resonances.
\end{abstract}


\maketitle % title page is now complete
\emph{Introduction.} Scattering properties in one-dimensional non-Hermitian potentials have received considerable attention recently[1-9]. It has been shown that $\mathcal{PT}$-symmetric quantum systems may exhibit symmetry-breaking transitions, from a real to a complex eigenvalues.[6-9], as the control parameter crossing the critical values. Due to the equivalent structure between the time-dependent Schrodinger equation and the paraxial electromagnetic wave equation, some remarkable experimental realizations of such $\mathcal{PT}$-symmetric transitions have been reported[2-5]. In one-dimensional $\mathcal{PT}$-symmetric photonic heterostructures many appealing scattering features have been found, such as the anisotropic transmission resonances, the coherent perfect absorber laser, and unidirectional invisibility[8-11]. More recently the effects of quantum interference on the critical behavior of $\mathcal{PT}$-symmetric potentials has been investigated in the context of tight-binding theory, where interesting phenomena like unidirectional perfect absorber, have been reported, based on the Aharonov-Bohm interferometer[12].

It is obvious that much richer quantum-interference induced scattering phenomena may be found in coupled $\mathcal{PT}$-symmetric systems, particularly in parallel connected potentials. For serially coupled one-dimensional, the transfer matrix provides a powerful means to study the transport properties[13-15]. It has been shown that in a one-dimensional potential consisting of $N$ identical cells the transmission and reflection amplitudes of the $N$-cell coupled system can be expressed in terms of the single-cell amplitudes and the Bloch phase, in a quite transparent manner[15]. On the other hand, for parallel connected systems, the scattering matrix method is usually employed for the two parallel coupled scatterers, as is well illustrated in the so-called Aharonov-Bohm quantum ring[16-20]. The tight-binding theory is another important approach, where the continuous physical space is approximated by discrete, connected sites. By its design this method provides a remarkable flexibility when used to study quantum transport on complex networks[21,22].

The purposes of this Letter is first to derive a general transparent $2\times 2$ transfer matrix formalism for quantum transport of $N$ one-dimensional arbitrary scattering cells, connected in parallel. This approach allows a transparent, closed-form expression of the transmission and reflection amplitudes as a function of the single-cell parameters. And secondly we apply our theory, together with the existing the transfer matrix formalism for one-dimensional serially coupled systems, to the studies of coupled $\mathcal{PT}$-symmetric scatterers. For convenience of demonstration we focus only on the $N$ identical $\mathcal{PT}$-symmetric potential, and give a general result without going into the details of model systems. We show that the coupled $\mathcal{PT}$-symmetric system may exhibit the characteristic spectral singularity and phase transition at the same or distinct critical points of the single-system control parameters,subject to the coupling configurations. In general, the similar transition patterns are observed para other parameter values, determined by both $\mathcal{PT}$-symmetric scattering potentials as well as the coupling mechanism. As a final conclusion we emphasize that our formalism may be regarded as recurrent relation if the individual scattering cell is considered as one level less scattering compound, and thus may apply to scalable scattering systems, like the glued N-ary Cayley trees[23].

\emph{Transfer matrix of parallel connection.} To better illustrate our idea we consider $N$ physical transport channels that all their left ends are joined together  at the joint site $O$ and all their right ends at the merging point $O'$. We denote by $\psi$ the wave function of the one dimensional Schrodinger equation, on the left lead, and $\phi$ the wave function on the right lead. More specifically, we have

\begin{equation}
\psi=Ae^{ikx}+Be^{-ik'x}\psi(x),\quad \phi=Ce^{iqx}+De^{-iq'x} \nonumber
\end{equation}
and on each branch we define
\begin{equation}
\psi_j=A_je^{ik_jx}+B_je^{-ik'_jx}, \quad \phi_j=C_je^{iq_jx}+De^{-iq'_jx}.\nonumber
\end{equation}
Here we have include the position-dependent effective mass which is assumed to be piecewise constant on each branch. We denote the mass on the $j$-th branch by $m_j$. The derivatives of the wave functions are given by
\begin{equation}
\frac{1}{m_j}\psi'_j=\frac{ik_j}{m_j}A_je^{ik_jx}-\frac{ik'_j}{m_j}B_je^{-ik'_jx}, \nonumber
\end{equation}
For convenience we further assume that
\begin{equation}
u_j=A_je^{ik_jx}, \quad  v_j=B_je^{-ik'_jx} \nonumber
\end{equation}
and
\begin{equation}
u'_j=C_je^{iq_jx}, \quad  v'_j=D_je^{-iq'_jx}, \nonumber
\end{equation}
with the above definitions the generalized Robin boundary condition at the vertex $O$ can be written in terms of $u_j$ and $v_j$, as follows
\begin{equation}
u_i+v_i=u_j+v_j=u+v, \quad i,j=1,2,...,N
\end{equation}
\begin{equation}
\alpha u-\beta v=\sum_{j=1}^{N}(\alpha_ju_j-\beta_jv_j)+\gamma (u+v),
\end{equation}
where
\begin{equation}
\gamma=-i\frac{2V_0}{\hbar^2}, \quad \alpha=\frac{k}{m}, \quad \beta=\frac{k'}{m} \nonumber
\end{equation}
\begin{equation}
\alpha_j=\frac{k_j}{m_j}, \quad \beta_j=\frac{k'_j}{m_j} \nonumber
\end{equation}

From Eqs.(1) and (2) it follows
\begin{equation}
\begin{bmatrix}
	u\\
	v \\
\end{bmatrix}
=\sum_{j=1}^{N}Q_j
\begin{bmatrix}
	u_j \\
	v_j \\
\end{bmatrix},
\end{equation}
where
\begin{equation}
Q_j=\frac{1}{N(\alpha + \beta)}
\begin{pmatrix}
	 \beta - \gamma+N\alpha_j & \beta - \gamma-N\beta_j  \\
	 \alpha + \gamma-N\alpha_j & \alpha+\gamma+N\beta_j \\
\end{pmatrix}
\end{equation}
Similarly, we have, at the contact point $O'$,
\begin{equation}
\begin{bmatrix}
	u'\\
	v' \\
\end{bmatrix}
=\sum_{j=1}^{N}Q'_j
\begin{bmatrix}
	u'_j \\
	v'_j \\
\end{bmatrix},
\end{equation}
with
\begin{equation}
Q'_j=\frac{1}{N(\alpha'+\beta')}
\begin{pmatrix}
	 \beta'+\gamma'+N\alpha'_j & \beta'+\gamma-N\beta'_j \\
	 \alpha'-\gamma'-N\alpha'_j & \alpha'-\gamma'+N\beta'_j \\
\end{pmatrix}
\end{equation}
where
\begin{equation}
\gamma'=-i\frac{2V'_0}{\hbar^2}, \quad \alpha'=\frac{q}{m'}, \quad \beta'=\frac{q'}{m'} \nonumber
\end{equation}
\begin{equation}
\alpha'_j=\frac{q_j}{m'_j}, \quad \beta'_j=\frac{q'_j}{m'_j} \nonumber
\end{equation}

To establish the relations between wave functions on different channels, we make us of the continuity condition of wave functions  at $O$ and $O'$, which read
\begin{equation}
u_i+v_i=u_j+v_j, \quad i,j=1,2,...,N
\end{equation}
\begin{equation}
u'_i+v'_i=u'_j+v'_j, \quad i,j=1,2,...,N
\end{equation}
Note that on the $i$-th branch, the wave functions at two joining points are related by
\begin{equation}
\begin{bmatrix}
	u_i \\
	v_i \\
\end{bmatrix}
=M_i
\begin{bmatrix}
	u'_i \\
	v'_i \\
\end{bmatrix}
=
\begin{pmatrix}
	 m^i_{1,1} & m^i_{1,2} \\
	m^i_{2,1} & m^i_{2,2} \\
\end{pmatrix}
\begin{bmatrix}
	u'_i \\
	v'_i \\
\end{bmatrix},
\end{equation}
For standard transfer matrix we have $ D_i=Det(M_i)=1$, but in general the inverse of $M_i$ can be written as
\begin{equation}
M^{-1}_i=\frac{1}{D_i}
\begin{pmatrix}
	 m^i_{2,2} & -m^i_{1,2} \\
	-m^i_{2,1} & m^i_{1,1} \\
\end{pmatrix},\nonumber
\end{equation}
Using the transfer matrices of the single scatterers we can express the wave functions $u'_i$ and $v'_i$ in terms of $u_i$ and $v_i$ in Eqs(7) and (8). Thus we obtain the relation that links the amplitudes on the i-th and j-th channel,
\begin{equation}
\begin{bmatrix}
	u_i \\
	v_i \\
\end{bmatrix}
=L_{i,j}
\begin{bmatrix}
	u_j \\
	v_j \\
\end{bmatrix}
\end{equation}
where
\begin{equation}
L_{i,j}=\frac{1}{D_i}\frac{1}{m^i_{1,1}-m^i_{1,2}+m^i_{2,1}-m^i_{2,2}}
\begin{pmatrix}
	 m^i_{1,1}-m^i_{1,2}+m^j_{2,1}-m^j_{2,2} & m^i_{1,1}-m^i_{1,2}-m^j_{1,1}+m^j_{1,2}\\
	 -m^i_{2,2}+m^i_{2,1}+m^j_{2,2}-m^j_{2,1} & -m^i_{2,2}+m^i_{2,1}+m^j_{1,1}-m^j_{1,2} \\
\end{pmatrix} \nonumber
\end{equation}
Substituting Eq.(10) into (3) we obtain the following relation which connects the wave function on the left lead to the amplitudes of an arbitrary channel $j$, through the point $O$,
\begin{equation}
\begin{bmatrix}
	u\\
	v\\
\end{bmatrix}
=\sum_{i=1}^{N}Q_iL_{i,j}
\begin{bmatrix}
	u_j \\
	v_j \\
\end{bmatrix}
\end{equation}

To emphasize the $N$ to $1$ reduction onto the effective single-channel transport, let us choose a specific channel as our reference path, for example, $j=s$, then we have
\begin{equation}
\begin{bmatrix}
	u\\
	v\\
\end{bmatrix}
=T_s
\begin{bmatrix}
	u_s \\
	v_s\\
\end{bmatrix}
\end{equation}
where
\begin{equation}
T_s=\sum_{j=1}^{N}Q_jL_{j,s}.
\end{equation}

Following the same procedure, one finds,
\begin{equation}
\begin{bmatrix}
	u'\\
	v'\\
\end{bmatrix}
=T'_s
\begin{bmatrix}
	u'_s \\
	v'_s\\
\end{bmatrix}
\end{equation}
with
\begin{equation}
T'_s=\sum_{j=1}^{N}Q'_jL'_{j,s}.
\end{equation}

Put together Eqs.(13)-(15) we obtain the transfer matrix for an array of scatterers coupled in parallel,
\begin{equation}
M =T_sM_sT'^{-1}_s
\end{equation}
Eq.(16) is our main result. Here the matrices $T_s$ and $T'_s$ represent the effects of the junctions $O$ and $O'$, determined by the coupling structure and the transfer matrices of all single scattering cells in between the two joining points. It is an $2 \times 2$ matrix equation with a transparent explanation, and much readily dealt with in both analytical and numerical calculations. This result is derived for a variety of junction potentials and scattering modes of individual one-dimensional localized potentials that are connected in parallel. Though an analytical solution can be achieved, the detailed computation may, in some cases, be rather lengthy and tedious, for arbitrary single scattering cells. As a straightforward illustration we re-consider the calculation of the transmission and reflection amplitudes for Aharonov-Bohm ring. With the notions used in Ref.[20], we have

\begin{equation}
k=\sqrt{\frac{2mE}{\hbar^2}}, \quad k_1=k+\frac{e\Phi}{\hbar cL}, \quad k_2=k-\frac{e\Phi}{\hbar cL}. \nonumber
\end{equation}

Here $\Phi$ is the magnetic flux and $L$ is the ring round length. Assume that $N=2$, $\alpha=\alpha'=0$, and $L=L_1+L_2$. After some algebra we obtain,
\begin{equation}
M_{2,2}=\frac{\Delta_1}{8(e^{ik_2L_1}+e^{ik_1L_2}-e^{-ik_1L_1}-e^{-ikL_2L_2})}
\end{equation}

\begin{equation}
M_{2,1}=\frac{\Delta_2}{8(e^{ik_2L_1}+e^{ik_1L_2}-e^{-ik_1L_1}-e^{-ikL_2L_2})}
\end{equation}
where
\begin{equation}
\Delta_1=-9e^{ik_2L_1+ik_1L_2}-e^{-ik_1L_1-ikL_2L_2}+4e^{-ik_2L_2+ik_1L_2}+4e^{-ik_1L_1+ik_2L_1}+e^{-ik_2L_2+ik_2L_1}+e^{-ik_1L_1+ik_1L_2} \nonumber
\end{equation}

\begin{equation}
\Delta_2=3e^{-ik_1L_1-ikL_2L_2}+3e^{ik_1L_2+ikL_2L_1}-4e^{-ik_1L_1+ik_2L_1}-4e^{-ik_2L_2+ik_1L_2}+e^{-ik_1L_1+ik_1L_2}+e^{-ik_2L_2+ik_2L_1} \nonumber
\end{equation}
The above results are slightly different in transmission and reflection amplitudes than those in Ref.[20], but under the condition $L_1=L_2=L/2$, we obtain the same transmittance:
\begin{equation}
|T|^2=(\frac{1}{M_{2,2}})^2=\frac{64}{\Delta^2}(1-\cos(kL))(1+cos(\psi)),
\end{equation}
with $\psi=-e\Phi/\hbar c$.

\emph{Coupled $PT$-symmetric scatterers.} Now we analyse the quantum transport of $N$ identical $PT$-symmetric scatterers connected in parallel. In this case, the identical scatterers are characterized by same single channel transfer matrices $M_i = M_j$, which yields $L_{i,j}=L'_{i,j}=I$. We assume that the contact potentials are vanishing at the the vertices $O$ and $O'$, considering them as the simple splitting and converting tunnels. For the same wave vectors on the two external leads and on each of the internal channels, that is, $\beta_j=\beta=\beta'_j=\beta'$, we obtain
\begin{equation}
T_s=T'_s=\frac{1}{2}
\begin{pmatrix}
	 N+1 & -(N-1) \\
	-(N-1) & N+1 \\
\end{pmatrix}
\end{equation}
As an interesting example we consider transport properties of one-dimensional $\mathcal{PT}$-symmetric localized potentials. Using the the following parameterized transfer matrix,
\begin{equation}
m=
\begin{pmatrix}
      a^* & ib \\
	  -ic & a  \\
\end{pmatrix},
\end{equation}
where $m_{1,1}=a*$, $m_{1,2}=-ib$, $m_{2,1}=ic$ and $m_{2,2}=a$. The $\mathcal{PT}$-symmetry imposes the constraints on the transfer matrix: $m_{2,2}(\omega)=m^*_{1,1}(\omega^{*})$, $m_{1,2}(\omega)=-m^*_{1,2}(\omega^{*})$ and $m_{2,1}(\omega)=-m^*_{2,1}(\omega^{*})$. The scattering matrix related to the transfer matrix (21) is given by
\begin{equation}
s=\frac{1}{a}
\begin{pmatrix}
      ib & 1 \\
	  1 & ic  \\
\end{pmatrix},
\end{equation}
Denoted by $\lambda_{\pm}$ the eigenvalues of the $S$ matrix, it follows that
\begin{equation}
\lambda_{\pm}=\frac{i}{2a}[(b+c) \pm \sqrt{(b-c)^2-4}] \nonumber
\end{equation}
From the eigenvalues of the $S$ matrix one finds that $\lambda_{+}\lambda_{-}=-|a|^2/a^2$. The $\mathcal{PT}$ symmetry is broken whenever $|\lambda_{+}|/|\lambda_{-}| > 1$. The exceptional points are defined by $b-c=\pm 2$, at which the eigenvalues of the $S$ matrix bifurcate. The spectral singularity is signatured by $m_{2,2}=0$. As pointed in [8,9], in optical medium the zeros of $m_{2,2}$ may be related to coherent perfect absorber (CPA) or $\mathcal{PT}$ laser, subject to the characteristic of the injected fields.

Inserting Eqs.(20) and (21) into (16) we obtain the transfer matrix of the whole parallel connected system. Its matrix elements are given by

\begin{eqnarray}
M_{1,1} & = &  \frac{1}{4N}[(N+1)^2a^*+i(N^2-1)(b+c)-(N-1)^2a] \nonumber \\
M_{1,2} & = &  \frac{1}{4N}[(N^2-1)a^*+i[(N+1)^2b+(N-1)^2c]-(N^2-1)a] \nonumber\\
M_{2,1} & = &  \frac{1}{4N}[-(N^2-1)a^*-i[(N-1)^2b+(N+1)^2c]+(N^2-1)a] \\
M_{2,2} & = &  \frac{1}{4N}[-(N-1)^2a^*-i(N^2-1)(b+c)+(N+1)^2a] \nonumber
\end{eqnarray}

It is readily verified that $\det(M)=1$. Introducing $M_{1,1}=a_N^*$, $M_{1,2}=-ib_N$, $M_{2,1}=ic_N$ and $M_{2,2}=a_N$ we find that $b_N-c_N=-(b-c)$. Therefore the exceptional points occurs at the same parameter value for both single and parallel coupled systems, indicating the resonance properties of the single scatterer is translated to the coupled system. This finding reveals that the $\mathcal{PT}$ symmetry is shared by the single scatterers and their parallel coupled compound, and the $\mathcal{PT}$ symmetry breaking is related only to the $\mathcal{PT}$-symmetric features of scattering potentials, as will be confirmed later, by serially coupled $\mathcal{PT}$-symmetric potentials.

Now let us examine the spectral singularities, which are given by $M_{2,2}(\omega_c)=0$ (the threshold for self-oscillation laser) and $M_{1,1}(\omega_c)=0$ (the perfect absorber). Here $\omega$ is the control parameter of the $\mathcal{PT}$-symmetric potential, and $\omega_c$ represents the critical value of the control parameter for $PT$ symmetry breaking. From (23) it follows that the critical value of the control parameter is different than that of the single scattering cell, the $\mathcal{PT}$-symmetry breaking of the coupled system occurs at some $\omega=\omega_N\neq \omega_c$, such that
\begin{equation}
\frac{N+1}{N-1}=\frac{i}{2a}[(b+c)\pm \sqrt{(b-c)^2-4}]
\end{equation}
This result shows that the broken $\mathcal{PT}$-symmetric phase may be induced by parallel coupling mode, in otherwise $\mathcal{PT}$-symmetric regimes of the single systems.

Now we look at the serially coupled $\mathcal{PT}$-symmetric scatterers. The transfer matrix $M$ of $N$ identical scattering cells, coupled in series, is described in terms of the single-cell transfer matrix $m$ by
\begin{equation}
M=\frac{1}{\sin(\phi)}[m\sin(N\phi)-\sin((N-1)\phi)]
\end{equation}
where the Bloch phase $\phi$ is given by $Tr(m)=2\cos(\phi)$. First we note immediately that $M_{1,2}=m_{1,2}\sin(N\phi)/\sin(\phi)$ and $M_{2,1}=m_{2,1}\sin(N\phi)/\sin(\phi)$, so that the exceptional points of the coupled system are given by $b-c=\pm 2\sin(\phi)/\sin(N\phi)$. Since the Block phase $\phi$ is related to the details of the scattering potential, the parameter critical value for $\mathcal{PT}$ symmetry breaking transition of the coupled scatterers is different than that of the individual one. It otherwise indicates that the single scattering cells may live in different phases in comparison with their coupled counterpart. Now let us analyse the spectral singularity of serially coupled $\mathcal{PT}$-symmetric potentials. From Eq.(25)it follows that
%\begin{equation}
%M_{1,1}=m_{1,1}\frac{\sin(N\phi)}{\sin(\phi)}-\frac{\sin((N-1)\phi)}{\sin(\phi)}, \quad %M_{2,2}=m_{2,2}\frac{\sin(N\phi)}{\sin(\phi)}-\frac{\sin((N-1)\phi)}{\sin(\phi)}
%\end{equation}
%\begin{equation}
%M_{1,2}=m_{1,2}\frac{\sin(N\phi)}{\sin(\phi)}, \quad M_{2,1}=m_{2,1}\frac{\sin(N\phi)}{\sin(\phi)}
%\end{equation}
%It can be seen that 
the singularities resulting from $M_{2,2}=0$ is given by $m_{2,2}=\sin((N-1)\phi)/\sin(N\phi)$. If the Bloch phase is chosen to be $\phi=\pi/(N-1)$, the spectral singularities and the $\mathcal{PT}$ symmetry breaking transition occur at the same parameter value for both the single cells and the serially connected such scattering units.

To be more specific, let us study the scattering properties of coupled $\mathcal{PT}$-synthetic Bragg scatterers[9]. The $\mathcal{PT}$-symmetric Bragg structure is defined by the refractive index distribution $n(z)=n_0+n_1\cos(2\beta z)+in_2\sin(2\beta z)$ for $|z|<L/2$. Here $\beta$ is the grating number, $k$ is the free propagation wave number. That is, it is assumed that the electric field is expressed by $E(z)=E_fe^{ikz}+E_be^{-ikz}$, outside the scattering region, while $E(z)=E_fe^{i(\delta+k)z}+E_be^{-i(\delta+k)z}$, inside the Bragg structure. $\delta=\beta-k$ is the detuning. Near the Bragg point, i.e., $\delta \approx 0$, the transfer matrix can be written as[Lin]
\begin{equation}
m=\frac{1}{\lambda}
\begin{pmatrix}
      \lambda\cos(\lambda L)-i\delta\sin(\lambda L) & i\frac{n_1+n_2}{2n_0}k\sin(\lambda L)  \\
	  -i\frac{n_1-n_2}{2n_0}k\sin(\lambda L) & \lambda\cos(\lambda L)+i\delta\sin(\lambda L)  \\
\end{pmatrix},
\end{equation}
As reported in Ref.[9], the Bragg structure exhibits a spontaneous $\mathcal{PT}$-symmetry breaking transition and unidirectional invisibility at $n_1=n_2$, corresponding to $m_{2,1}=0$ in the description of transfer matrix. It is interesting to see that, for $N$ $\mathcal{PT}$-symmetric Bragg scatterers connected in parallel, $M_{2,1}=0$ results in
\begin{equation}
n_2=\frac{N^2-1}{N}\frac{n_0\delta }{k}+\frac{N^2+1}{2N}n_1
\end{equation}
while for $n_1=n_2$, it reduces to that $n_2=2n_0\delta(N+1)/k(N-1)$. This shows that the exceptional points are modified by the parallel coupling. In the case of $n_1=n_2$, the coupling imposes additional constraints $n_1=n_2=2n_0\delta(N+1)/k(N-1)$. On the other hand, there exist new possibilities for $\mathcal{PT}$-symmetry breaking transition, induced by the parallel connections, as is indicated by (29). More detailed discussion will be presented in a separate work.

We turn to the case of the serial connection, where the relations between the transfer matrix of the coupled structures and those of the single units are given by (26) and (27). It is remarkable that the Bloch phase $\phi=\lambda L+2n\pi$ with $n=0, \pm1,\pm2, ...$, as a result of $Tr(m)=2\cos(\phi)$. From (27) it follows immediately that $M_{2,1}=0$ whenever $m_{2,1}=0$.This implies that all exceptional points of the single Bragg structures are also the phase transition points for the coupled one. In addition, new spontaneous $\mathcal{PT}$-symmetry breaking transition occurs at $\phi = m\pi/N$ for $m=0, \pm 1, \pm 2, ...$, which is well-known characteristic of one-dimensional periodic potentials.

\emph{Conclusion.} We derive an transfer matrix approach for quantum transport through parallel connected one-dimensional scattering cells. An exact, closed-form expression for the transfer matrix of $N$ identical scatterers in parallel is given. We report some novel scattering properties of coupled $\mathcal{PT}$-symmetric potentials. It is shown that the spectral singularities and $\mathcal{PT}$-symmetry breaking transition may be induced by different coupling modes. Those coupling induced features may result from the individual constituent single scatterers, as in the case of serial connection patterns, and may be at totally distinct parameter values than the critical regime of the scattering units. It is worthwhile pointing out that as a recurrence relation, our transfer matrix formalism can be readily applied to the quantum transport on a glued n-ary Carley trees, and other scalable structures of scattering objects.


\begin{thebibliography}{99}

\bibitem{Bender} C. M. Bender and S. Boettcher, Phys. Rev. Lett. 80, 5243 (1998).

\bibitem{Regens} A. Regensburger, C. Bersch, M.-A. Miri, G. Onishchukov, D.N. Christodoulides, and U. Peschel, Nature (London) 488, 167 (2012).

\bibitem{Makris} K. G. Makris, R. El-Ganainy, D. N. Christodoulides, and Z. H. Musslimani, Phys. Rev. Lett. 100, 103904 (2008).

\bibitem{Guo} A. Guo, G. J. Salamo, D. Duchesne, R.Morandotti, M. Volatier-Ravat, V. Aimez, G. A. Siviloglou, and D. N. Christodoulides, Phys. Rev. Lett. 103, 093902 (2009).

\bibitem{Rüter} C. E. R$\ddot{u}$ter, K.G.Makris, R. El-Ganainy,D.N. Christodoulides, M. Segev, and D. Kip, Nat. Phys. 6, 192 (2010).

\bibitem{Chong} Y. D. Chong, L. Ge, and A. D. Stone, Phys. Rev. Lett. 106, 093902 (2011).

\bibitem{Ge} L. Ge, Y. D. Chong, and A. D. Stone, Phys. Rev. A 85, 023802 (2012).

\bibitem{Longhi} S. Longhi, Phys. Rev. A 82, 031801(R) (2010).

\bibitem{Lin} Zin Lin, Hamidreza Ramezani, Toni Eichelkraut, Tsampikos Kottos, Hui Cao, and Demetrios N. Christodoulides Phys. Rev. Lett. 106, 213901 (2011).

\bibitem{Mostafa} A. Mostafazadeh, Phys. Rev. A 80, 032711 (2009).

\bibitem{Ramezani} Hamidreza Ramezani, Hao-Kun Li, Yuan Wang, and Xiang Zhang, Phys. Rev. Lett. 113, 263905 (2014).

\bibitem{Song} L. Jin, P. Wang and Z. Song, Scientific Report, 6 32919 (2016)

\bibitem{Griff}D.J. Griffiths and C. A. Steinke, Am. J Phys. 69 137 (2001).

\bibitem{CH Wu}C.H. Wu and G. Mahler, Phys. Rev. B 43 5012 (1991).

\bibitem{Sprung}D.W.L. Sprung and H. Wu, Am. J Phys. 61 1118 (1993).

\bibitem{Shapiro} B. Shapiro, Phys. Rev. Lett. 50 747 (1983).

\bibitem{Buttiker} M. B$\ddot{u}$ttiker, Y. Imry, and M. Ya. Azbel, Phys. Rev. A 30 1982 (1984).

\bibitem{Gefen}Y. Gefen, Y. Imry, and M.Y. Azbel, Phys. Rev. Lett. 52 129 (1984).

\bibitem{Cahay} M. Cahay, S. Bandyopadhyay, and H. L. Grubin, Phys. Rev. B 39 12989 (1989).

\bibitem{Xia} J.B. Xia, Phys. Rev. B 45 3593 (1992).

\bibitem{Kostrykin} V. Kostrykin and R. Schrader, J. Phys. A 32 595 (1999)

\bibitem{Andradea}F. M. Andradea, A.G.M. Schmidtc, E. Vicentinid, B.K. Chenge, M.G.E. da Luze, Physics Reports, 647 1 (2016).

\bibitem{Yu} Yu Jiang, M. Martínez-Mares, E. Castano, and A. Robledo, Phys. Rev. E 85, 057202 (2012).
\end{thebibliography}
\end{document}